\documentclass[preprint,trackchanges]{JASA}

\usepackage{tipa}
\usepackage[utf8]{inputenc}
\usepackage{amssymb}
\usepackage{multirow}
\graphicspath{{./Figures/}}

\begin{document}
\rightlinenumbers*

\title[Triangular body-cover model with intrinsic muscle control]{
Triangular body-cover model of the vocal folds with coordinated activation of the five intrinsic laryngeal muscles
        }



\author{Gabriel A. Alzamendi}
\email{galzamendi@ingenieria.uner.edu.ar}
\affiliation{
Institute for Research and Development on Bioengineering and Bioinformatics (IBB), CONICET-UNER, Oro Verde, Entre Ríos 3100, Argentina}
\altaffiliation{Also at: Facultad de Ingeniería, Universidad Nacional de Entre Ríos,  Oro Verde, Entre Ríos 3100, Argentina.}
\author{Sean D. Peterson}
\email{peterson@uwaterloo.ca}
\affiliation{Mechanical and Mechatronics Engineering, 
University of Waterloo, Waterloo, Ontario N2L 3G1, Canada}
\author{Byron D. Erath}
\email{berath@clarkson.edu}
\affiliation{Department of Mechanical and Aeronautical Engineering, 
Clarkson University, Potsdam, New York 13699, USA}
\author{Robert E. Hillman}
\email{hillman.robert@mgh.harvard.edu}
\affiliation{Center for Laryngeal Surgery and Voice Rehabilitation, Massachusetts General Hospital, Boston,Massachusetts 02114, USA}
\altaffiliation{Also at: Department of Surgery, Massachusetts General Hospital–Harvard  Medical  School, Boston, MA02115, USA.}
\author{Matías Zañartu}
\email{matias.zanartu@usm.cl}
\affiliation{Department of Electronic Engineering, 
Universidad Técnica Federico Santa María, Valparaíso 2390123, Chile}

\date{\today} 

\begin{abstract}
Poor laryngeal muscle coordination that results in abnormal glottal posturing is believed to be a primary etiologic factor in common voice disorders such as non-phonotraumatic vocal hyperfunction. Abnormal activity of antagonistic laryngeal muscles is hypothesized to play a key role in the alteration of normal vocal fold biomechanics that results in the dysphonia associated with such disorders. Current low-order models of the vocal folds are unsatisfactory to test this hypothesis since they do not capture the co-contraction of antagonist laryngeal muscle pairs. To address this limitation, a self-sustained triangular body-cover model with full intrinsic muscle control is introduced. The proposed scheme shows good agreement with prior studies using finite element models, excised larynges, and clinical studies in sustained and time-varying vocal gestures. Simulations of vocal fold posturing obtained with distinct antagonistic muscle activation yield clear differences in kinematic, aerodynamic and acoustic measures. The proposed tool is deemed sufficiently accurate and flexible for future comprehensive investigations of non-phonotraumatic vocal hyperfunction and other laryngeal motor control disorders. 
\end{abstract}

\maketitle

\section{Introduction} \label{SecI}

Non-phonotraumatic vocal hyperfunction (NPVH) is a common voice disorder associated with excessive and poorly regulated activity of the intrinsic and extrinsic laryngeal muscles \citep{Hillman_framework_2020}, causing a range of different types of disordered voice quality \citep{vanstan_differences_2021} but without trauma to the vocal fold (VF) tissue. A common manifestation of NPVH during phonation is high levels of stiffness and tension in the VFs accompanied by incomplete glottal closure causing the voice to be rough, strained, and breathy with increased subglottal pressure and slightly higher, less periodic, and less variable pitch \citep{ Hillman_framework_2020, Espinoza2017, Espinoza2020,  vanstan_differences_2021}. Very little is known about the specific  physical mechanisms that underlie this or other manifestations of NPVH. However, the apparent increase in muscle activity and vocal fold stiffness, abducted glottal configuration, and relatively restricted pitch variability suggest that poor coordination of antagonist laryngeal muscle pairs plays a key role in the altered biomechanics and resulting dysphonia associated with  NPVH.

Pitch, loudness and quality of the voice are primarily controlled by laryngeal muscles in a complex process that is still not fully understood. The process involves the coordinated action of intrinsic and extrinsic muscles  \citep{huber_control_2004}, neural muscle effects \citep{titze_reflex_2002}, and auditory and somatosensory feedback and feedforward mechanisms \citep{Lester2020}. Though no mathematical model currently captures all of these factors, efforts have been made to describe components of laryngeal muscle control by either simplified biomechanical representations \citep{ farley_biomechanical_1996,titze_rules_2002, titze_twodimensional_2007} or high-fidelity three-dimensional finite element models \citep{ geng_threedimensional_2020, gommel_muscle_2007, hunter_threedimensional_2004, movahhedi_effects_2021,
yin_interaction_2014}. These high order models can capture the complex biomechanical and geometrical changes due to muscle activation but are likely too computationally demanding to account for neural motor control effects. Low-order models can better handle the latter task \citep{manriquez_neurophysiological_2019} and are more suitable for comprehensive parametric simulations that would be needed in the context of laryngeal motor control. However, there is currently no muscle activation framework for low-order lumped-element models of the vocal folds that incorporates the possibility of independent co-contraction of all agonist/antagonist intrinsic muscle pairs. 

The present study introduces a physiologically-based scheme for controlling the mechanical properties of a self-sustained, low-order model of the vocal folds through both independent and coordinated activation of all five intrinsic laryngeal muscles. The approach builds upon prior efforts that describe rules for controlling low-order models \citep{titze_rules_2002}, vocal fold posturing \citep{titze_twodimensional_2007}, and a triangular body-cover vocal fold model \citep{galindo_modeling_2017}. The scheme provides a flexible and physiologically relevant way to control the self-sustained, fully interactive voice production model for both sustained vowels and time-varying glottal gestures that enables the exploration of the role of antagonistic muscle pairs in phonation. In this study, we illustrate the capabilities of the proposed scheme and contrast the simulations against prior numerical and experimental studies. Future studies will be devoted to comprehensively exploring the range of vocal disturbances associated NPVH with the proposed scheme.

The paper is organized as follows. In \autoref{SecII}, the different components of the proposed model are introduced.  In \autoref{SecIII}, results are presented for posturing, steady and dynamic phonatory gestures, and antagonistic muscle behavior. In \autoref{Sec.Disc}, the discussion of the results is provided, along with the proposes guidelines for future work.  Finally, \autoref{Sec.Concl} summarizes the work and provides the main conclusions of the study.

\section{Physiologically-controlled voice production model} \label{SecII}
Herein, we aim to simulate the muscle control exerted by the coactivation of the five intrinsic laryngeal muscles during phonation: cricothyroid (CT), thyroarytenoid (TA), lateral cricoarytenoid (LCA), interarytenoid (IA), and posterior cricoarytenoid (PCA); along with the passive contribution of the vocal ligament (LIG) and VF mucosa (MUC). Laryngeal function is described herein in terms of the glottal posture, resulting from the accommodation of the laryngeal cartilages, and of the VF oscillations around the phonatory configuration. 

The main components of the proposed model are comprised of a VF posturing scheme, the triangular body-cover VF model, an extended set of physiological rules for the VF model, and a vocal tract model. Each component is introduced in this section and their interrelationships are described.

\subsection{Vocal fold posturing model} \label{SecIIA}
  
\begin{figure*}[t]
  \figline{\fig{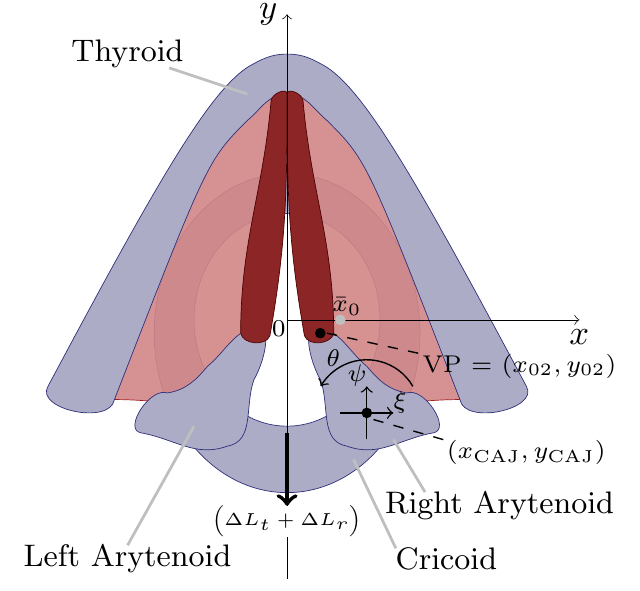}{8cm}{(a)}\label{FIG1a}
  \fig{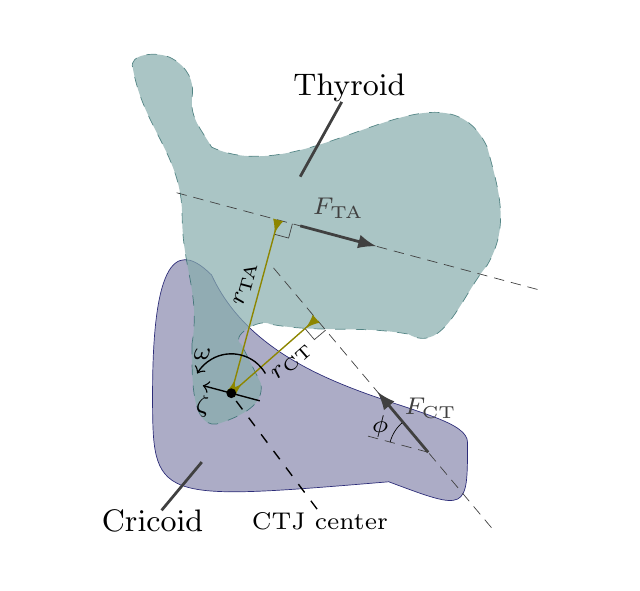}{8cm}{(b)}\label{FIG1b}}
  \caption{\label{FIG1} (Color online) Main laryngeal structures involved in the prephonatory posturing. (a) Cricoarytenoid accommodation at the glottal plane; and (b) Projection of effective cricothyroid accommodation on the glottal plane. Glottal geometry and vocal fold adjustment are controlled via the relative accommodation of major laryngeal cartilages. Figures adapted from \citet{titze_myoelastic_2006}. CAJ: cricoarytenoid joint, CTJ: cricothyroid joint.}
\end{figure*}

Phonatory posturing refers to the large, slowly-varying changes (relative to VF oscillations) in the glottal geometry due to the accommodation of the surrounding structures and the mechanics of the laryngeal tissues, which can be described in terms of the relative movement of the main laryngeal cartilages (see \autoref{FIG1}). The cricoid cartilage is a ring-shaped structure that delimits the larynx inferiorly. It provides support anteriorly to the thyroid cartilage, a large shield-like structure forming the main anterior framework of the larynx, and posteriorly to the paired small pyramidal-shaped arytenoid cartilages. The anterior angle in each arytenoid, referred to as the vocal process (VP), serves as the posterior points of attachment for the VFs. Adduction/abduction comes from the movement of the arytenoid cartilages around the cricoarytenoid joint (CAJ), the anatomical link with the cricoid cartilage (see \autoref{FIG1a}). Whereas adduction is controlled by the LCA, IA, and TA muscles, abduction is mainly controlled by the PCA muscle with minor contribution from the CT muscle \citep{chhetri_neuromuscular_2012, chhetri_influence_2014, geng_threedimensional_2020, hunter_threedimensional_2004}. On the other hand, the relative movement of the thyroid and cricoid cartilages around the cricothyroid joint (CTJ), as shown in \autoref{FIG1b}, is driven by the co-activation of the CT and TA muscles and plays a key role in VF elongation
\citep{chhetri_influence_2014, geng_threedimensional_2020}. 

The biomechanical model introduced by \citet{titze_twodimensional_2007} was applied to capture the main effects of the laryngeal musculature on the laryngeal configuration.
A detailed description of the model development and the measurement of the biomechanical parameters can be found in Chapter 3 of  \citet{titze_myoelastic_2006}. The model is built around the relative configuration of laryngeal cartilages; namely, that of the arytenoid cartilages with respect to the cricoid cartilage at the CAJ, and of the thyroid cartilage with respect to the cricoid cartilage at the CTJ.
Phonatory configuration is described in a transverse plane at the glottal level (see \autoref{FIG1a}).
The global coordinate system is set with the $y$-axis along the midplane in the glottis, pointing anteriorly, and the $x$-axis passing through the vocal process in the cadaveric position, pointing to the right. Hence, the Cartesian coordinates of the right VP in the cadaveric state is $(\bar{x}_{0},\,0)$, and geometric symmetry across the midplane is henceforth assumed.

The complex movements involved in arytenoid accommodation are modeled as coordinated ``effective'' movements of translation and rotation relative to the fixed cadaveric CAJ center, $(x_\text{CAJ}, \, y_\text{CAJ})$.
Let $(\xi, \, \psi)$ and $\theta$ be, respectively, the relative Cartesian displacements and the angle of rotation of the right arytenoid cartilage with respect to the CAJ center, as described in \autoref{FIG1a}. The equations of motion are as follows \citep{titze_twodimensional_2007, titze_myoelastic_2006}:
\begin{subequations}
\begin{eqnarray}
  \label{Eq:S2s1_01}
  M_\text{ac} \, \ddot{\xi} + d_{x} \dot{\xi} + k_{x} \xi & = & \sum_{i \in \mathcal{I}} \alpha_{i} F_{i}, \\
  \label{Eq:S2s1_02}
  M_\text{ac} \, \ddot{\psi} + d_{y} \dot{\psi} + k_{y} \psi & = & \sum_{i \in \mathcal{I}} \beta_{i} F_{i}, \\
  \label{Eq:S2s1_03}
  I_\text{ac} \, \ddot{\theta} + \delta \dot{\theta} + \kappa \theta & = & \sum_{i \in \mathcal{I}} \gamma_{i} F_{i}, 
\end{eqnarray}
\end{subequations}
where a dot over a variable indicates differentiation with respect to time, $t$. $F_{i}$ denotes the point force magnitude for each element in the set of laryngeal tissues $\mathcal{I}$ = \{LCA, IA, PCA, CT, TA, LIG, MUC\}, and $\alpha_{i}$, $\beta_{i}$, and $\gamma_{i}$ are the associated directional cosines and directional moment, as reported by \citet{titze_twodimensional_2007}. Biomechanical parameters $M_\text{ac}$ and $I_\text{ac}$ are the mass and moment of inertia of the arytenoid cartilage; $k_{x}$, $k_{y}$, and $\kappa$  are the translational and rotational stiffnesses (computed according to Chapter 3 in \citet{titze_myoelastic_2006}); and $d_{x}$, $d_{y}$, and $\delta$ are the translational and rotational damping coefficients, respectively.

Following \citet{titze_myoelastic_2006}, the relative cricoid accommodation with respect to the thyroid cartilage is in turn simulated in terms of translational, $\zeta$, and rotational, $\omega$,  movements around the fixed CTJ center, as illustrated in \autoref{FIG1b}. A simple scheme for projecting the effects of CT muscle activation on the horizontal glottal plane is considered, which allows assessment of the concomitant changes on the glottal configuration and vocal fold elongation. It is based on the following second-order equations:
\begin{subequations}
\begin{eqnarray}
  \label{Eq:S2s1_04}
  M_\text{cc} \ddot{\zeta} + k_{t} \left( t_{t} \dot{\zeta} + \zeta \right) & = & \left[ \cos{\phi} \, F_\text{CT} - \left(F_\text{TA} + F_\text{LIG} + F_\text{MUC} \right) \right], \\
  \label{Eq:S2s1_05}
  I_\text{cc} \ddot{\omega} + k_{r} \left( t_{r} \dot{\omega} + \omega \right) & = & \left[ r_\text{CT} F_\text{CT} - r_\text{TA} \left(F_\text{TA} + F_\text{LIG} + F_\text{MUC} \right) \right],
\end{eqnarray}
\end{subequations}
where $M_\text{cc}$ and $I_\text{cc}$ are the mass and moment of inertia associated with translation and rotation of the cricoid-arytenoid complex around the CTJ, $k_{t}$ and $k_{r}$ are the translational and rotational stiffnesses (computed according to Chapter 3 in \citet{titze_myoelastic_2006}), $t_{t}$ and $t_{r}$ are the time constants for the translational and rotational viscous damping, $r_\text{TA}$ and $r_\text{CT}$ are the moment arms for the TA and CT torques relative to CTJ center, 
and $\phi$ is the angle between the line of action of $F_\text{CT}$ and the translation axis of the cricoid cartilage, respectively. 

Cricoid accommodation gives rise to a concomitant posterior displacement, $(\Delta L_{t} + \Delta L_{r})$, in the arytenoid cartilage, as illustrated in \autoref{FIG1a}, where $\Delta L_{t} = \zeta$ and $\Delta L_{r} = r_\text{TA} \, \omega$ are the resulting posture effects due to the translation and rotation around the CTJ, respectively. As the VP is structurally linked to the arytenoid cartilage, its Cartesian coordinates are obtained as follows \citep{titze_twodimensional_2007}:
\begin{subequations}
\begin{eqnarray}
  \label{Eq:S2s1_06}
  x_{02} & = & x_{CAJ} - (x_{CAJ} - \bar{x}_{0}) \cos{\theta} + y_{CAJ} \sin{\theta} + \xi, \\
  \label{Eq:S2s1_07}
  y_{02} & = & y_{CAJ} (1 - \cos{\theta}) - (x_{CAJ} - \bar{x}_{0}) \sin{\theta} + \psi - (\Delta L_{t} + \Delta L_{r}). 
\end{eqnarray}
\end{subequations}
The contribution to vocal fold elongation due to arytenoid adduction is obtained by  \citep{titze_twodimensional_2007}:
\begin{equation}
  \label{Eq:S2s1_08}
  \Delta L_{a} = - \left[ y_{CAJ} (1 - \cos{\theta}) - (x_{CAJ} - \bar{x}_{02}) \sin{\theta} + \psi \right].
\end{equation}
Then, the total VF elongation is modeled as the sum of the translational, rotational, and adductory components, and thus the vocal fold strain is given by:
\begin{equation}
  \label{Eq:S2s1_09}
   \epsilon = \frac{1}{L_{0}} \, \left(\Delta L_{a} + \Delta L_{t} + \Delta L_{r} \right),
\end{equation}
where $L_{0}$ is the cadaveric VF length. Equation \eqref{Eq:S2s1_09} dynamically couples the adduction and elongation procedures \citep{titze_myoelastic_2006, titze_twodimensional_2007}.  The elongated VF length is $L_{g} = (1+\epsilon) \, L_{0}$.

In order to simulate the internal stress-strain response in the laryngeal tissues and to obtain the forces $F_{i}$, $i \in \mathcal{I}$, the modified  one-dimensional Kelvin model of elongation  is applied \citep{hunter_threedimensional_2004, hunter_refinements_2007}. The modeled tissue force accounts for the active component due to the internal muscle contractile forces, and the passive response force in connective tissue.
The activation level for each intrinsic muscle is then controlled through a normalized coefficient, i.e., $0.0\leq a_{i} \leq 1.0$, for $i \in$ \{LCA, IA, PCA, CT, TA\}. Because the vocal ligament and mucosa are non-contractile connective tissues, the active component is set to zero for $i \in $\{LIG, MUC\}. In \autoref{Sec.KelvinModel}, the muscle model and the biomechanical parameters for the laryngeal tissues are described in more detail.

\subsection{Triangular body-cover model of the vocal folds} \label{SecIIB}
\begin{figure}[t]
  \includegraphics[width=8cm]{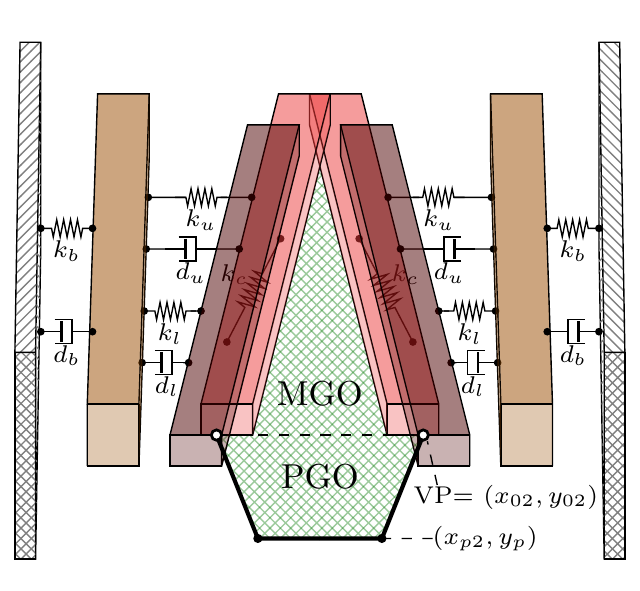}
  \caption{\label{FIG2} (Color online) Schematic of the triangular body-cover model of the vocal folds.
}
\end{figure}

The triangular body-cover model (TBCM) introduced by \citet{galindo_modeling_2017} is revised for numerically modeling the VF oscillations. \autoref{FIG2} provides a schematic of the TBCM model, with the associated tissue parameters and coordinates, where the subscripts $u$, $l$, and $b$ henceforth denote the upper, lower, and body masses, respectively. The VF model is based on the layered approximation proposed by \citet{story_voice_1995} for capturing the body-cover structure of the VFs, where two masses stacked in the inferior-superior direction model the cover and a single large mass situated laterally encapsulates the body dynamics. These lumped elements are coupled via non-linear mechanical elements to account for tissue viscoelasticity. In addition, the TBCM incorporates a gradual tilt of the VF edges in the anterior-posterior direction as a function of the degree of abduction, as is shown in \autoref{FIG2}, giving rise to a triangular glottis accounting for the anatomical dorsal-ventral gradient in the VFs, and the zipper-like closure commonly observed in female phonation \citep{birkholz_synthesis_2011, birkholz_articulatory_2011}. The TBCM formulation considers the structural link between the membranous area and the posterior glottal opening extended onto the cartilaginous glottis \citep{erath_review_2013, zanartu_modeling_2014}.

Originally, the TBCM was built around the simplification that muscle control only affects the VF adjustment, so the arytenoid accommodation was independently parameterized through \emph{ad hoc} rotation and displacement parameters. In this study, on the contrary, it is assumed that arytenoid accommodation is solved within the laryngeal posturing framework described in \autoref{SecIIA}. Hence, the adductory displacement in the posterior border of the upper mass, relative to the anterior commissure, coincides with the VP horizontal coordinates $x_{02}$ in Equation \eqref{Eq:S2s1_06} (see \autoref{FIG2}). For the adductory displacement of the lower mass, $x_{01}$, the net glottal convergence is also considered following the ideas in \citet{titze_rules_2002}. 

The coupled equations for simulating the TBCM dynamics are:
\begin{subequations}
  \label{Eq:S2s2_01}
 \begin{eqnarray}
  \label{Eq:S2s2_01a}
  F_{u} & = & m_{u} \, \ddot{x}_{u} = F_{k,\,u} + F_{d,\,u} - F_{kc} + F_{e,\,u} + F_{Col,\,u}, \\ 
  \label{Eq:S2s2_01b}
  F_{l} & = & m_{l} \, \ddot{x}_{l} = F_{k,\,l} + F_{d,\,l} + F_{kc} + F_{e,\,l} + F_{Col,\,l}, \\ 
  \label{Eq:S2s2_01c}
  F_{b} & = & m_{b} \, \ddot{x}_{b} = F_{k,\,b} + F_{d,\,b} - \left[F_{k,\,u} + F_{d,\,u} + F_{k,\,l} + F_{d,\,l} \right],
  \end{eqnarray}
\end{subequations}
where $m$ is mass, $x$ is the medial-lateral displacement over time, and $F$ is the force component for each block. Rest positions for the upper and lower blocks are $x_{u,\,0}= 0.5 \, x_{02}$, $x_{l,\,0}= 0.5 \, x_{01}$ pursuant to VF posturing, whereas $x_{b,\,0}= 3$ mm \citep{galindo_modeling_2017}. Furthermore, force subscripts $k$, $d$, $e$, and $kc$ represent the mechanical forces produced by the springs, dampers, flow pressures, and elastic coupling between the upper and lower masses, respectively. An additional spring force, $F_{Col}$, is introduced during vocal fold collision to capture the effects of impact between opposing upper/lower cover masses. The definitions of the forces are described in detail in the appendix of \citet{galindo_modeling_2017}, and are thus omitted here for brevity. To compute the aerodynamic driving forces, the intraglottal pressures on the upper/lower cover masses are derived from the subglottal pressure, $P_{s}$, and the supraglottal pressure, $P_{e}$, according to the formulation introduced in the appendix of \citet{titze_regulating_2002}.

\subsection{Physiological rules for the triangular body-cover model} \label{SecIIC}

The empirical rules introduced by \citet{titze_rules_2002} are applied and extended for the computation of the TBCM parameters. These rules introduce dynamic muscle control based on mapping the normalized activation levels of the intrinsic muscles, introduced at the end of \autoref{SecIIA}, into the relevant mechanical parameters (e.g., linear stiffness, mass distribution, glottal convergence, and VF length, thickness, and depth). Besides controlling the TBCM vibrations, the rules also have an impact on other aerodynamic and acoustic quantities in scenarios involving the tissue-flow-acoustic interaction in the glottis \citep{lowell_simulated_2006, zanartu_modeling_2014}. 

Originally, \citet{titze_rules_2002} begin by prescribing the VF elongation and VP adduction as a function of normalized activation levels $a_\text{TA}$, $a_\text{CT}$, and $a_\text{LC}$, with the latter combining the effects of both the LCA and PCA muscles. The geometrical parameters and fiber stresses are then obtained from the VF elongation, after which the stiffness and mass distributions in the VF model are computed. In this work, a few methodological modifications are considered to extend the rules for the TBCM. The first difference consists of including independent activation levels for the five intrinsic muscles; hence, the activation set $\mathbf{a} = [a_\text{LCA}, \,a_\text{IA}, \,a_\text{PCA}, \,a_\text{CT}, \,a_\text{TA}]$ controls the activation of the laryngeal musculature. Moreover, internal stresses in the VF tissues, specifically for the TA muscle, LIG, and MUC, are dynamically obtained by solving a modified Kelvin model (see \autoref{Sec.KelvinModel}), whereas the VF elongation and VP adduction are obtained from the laryngeal posturing according to Equations \eqref{Eq:S2s1_09} and \eqref{Eq:S2s1_06}, respectively; hence, the elongation and adduction rules in \citep{titze_rules_2002} are replaced. With this information, the remaining rules in \citet{titze_rules_2002} are thus applied for computing the nodal point, thickness, and depth for each block, and glottal convergence, from which the biomechanical parameters (i.e., mass and spring values) of the TBCM are derived.
Notably, the shear moduli of the body and cover are set to $\mu_{b} = 600$ Pa and $\mu_{c} = 300$, respectively, which differs from the values selected by \citet{titze_rules_2002}.

Based on VF adduction and symmetry with respect to the midsagittal plane, the glottal area for the upper and lower cover masses are:
\begin{subequations}
 \label{Eq:S2s3_01}
 \begin{eqnarray}
 \label{Eq:S2s3_01a}
 A_{u} & = & 2 (1 - \alpha_{u}) L_{g} \big(\tilde{x}_{u} + 0.5 (1 + \alpha_{u}) x_{01} \big), \\
 \label{Eq:S2s3_01b}
 A_{l} & = & 2 (1 - \alpha_{l}) L_{g} \big(\tilde{x}_{l} + 0.5 (1 + \alpha_{l}) x_{02} \big),
\end{eqnarray}
\end{subequations}
where $\tilde{x}_{u} = x_{u} - x_{u,\,0}$ and $\tilde{x}_{l} = x_{l} - x_{l,\,0}$ are block displacements relative to their rest positions. Additionally, $\alpha_{u}$ and $\alpha_{l}$ are the proportions of mass length for the upper and lower blocks undergoing collision at the given time, where $0.0 \leq \alpha_{u},\alpha_{l} \leq 1.0$. As a result, the area for the membranous glottal opening is $A_\mathrm{MGO} = \min\{A_{u}, \, A_{l}\}$.  
The effects of laryngeal posture on the posterior cartilaginous portion of the glottis are also simulated. Following \citet{titze_myoelastic_2006}, a trapezoidal shape is assumed for the posterior glottal opening, thus the resulting area can be computed as follows:
\begin{equation}
 \label{Eq:S2s3_02}
  A_\mathrm{PGO} = \max \{0, \, \min\{(x_{p1}+ x_{01}), \, (x_{p2}+ x_{02}) \} (y_{02} - y_{p}) \},
\end{equation}
where $x_{p1}$ is the posterior wall half-width at the bottom,  $x_{p2}$ is the posterior wall half-width at the top, and $y_{p}$ is the posterior wall position along the longitudinal axis.
The total glottal area comprises both the membranous and the cartilaginous parts:
\begin{equation}
  \label{Eq:S2s3_03}
  A_g = A_\mathrm{MGO} + A_\mathrm{PGO}.    
\end{equation}

\subsection{Interactions at the glottis and acoustic wave propagation} \label{SecIID}

To capture the physics of human phonation, the three-way interaction between sound, flow, and VF tissue is considered. For computing the air volume velocity, $U_g$, through the glottal area, $A_g$ (see Equation \eqref{Eq:S2s3_03}), the solver proposed in \citet{zanartu_modeling_2014} considering the acoustic driving pressures and the posterior glottal opening is applied, with the inclusion of the corrections made by \citet{lucero_smoothness_2015}. As shown in \citet{zanartu_modeling_2014}, in the one-dimensional flow approximation the solution obtained for a domain with two separate orifices for the posterior and membranous areas is equivalent to solving for the volume velocity through the total glottal area, $A_g$.

The wave reflection analog scheme is selected for describing the propagation of one-dimensional, planar acoustic waves in the time domain throughout the equivalent subglottal and supraglottal systems \citep{zanartu_influence_2006}. These tracts are discretized as the concatenation of a finite number of short uniform cylinders with variable cross-sectional areas. The area functions obtained from magnetic resonance imaging data for a male participant during sustained phonation are applied to simulate the supgraglottal tract for different vowels \citep{story_comparison_2008} and the subglottal system \citep{zanartu_influence_2006}. For simulating an equivalent excised-larynx scenario within the same framework, the non-interactive control scenario introduced by \citet{titze_sensitivity_2016} is also considered, where a very wide cross-sectional area (30 cm$^2$) is set for every cylinder in both the subglottal and the supraglottal systems. Boundary conditions based on a resistive lung termination and a parallel resistive-inertive element at the lips are also defined in the simulations. 

\subsection{Numerical implementation} \label{SecIIE}
The implemented model involving the muscle control of the larynx posture and vocal fold function depends upon several anatomical and biomechanical parameters. Model parameters were selected to simulate the physiology of male phonation. The parameters considered in this work are reported in \autoref{Tab1}. As in \citet{galindo_modeling_2017}, the simulations were performed using a truncated Taylor-series approximation to solve the differential equations. A sampling frequency of 44.1 kHz was employed. A colored random source component was included for modeling the turbulent aspiration noise generated at the glottis \citep{galindo_modeling_2017}.
 
\section{Results} \label{SecIII}
This section introduces illustrative simulations with the proposed voice production model. The effects of intrinsic muscle activation on the laryngeal posture are first described, then simulations for sustained vowels and articulatory gestures are subsequently analyzed. An example resembling antagonistic muscle behavior during sustained phonation is also discussed.

\subsection{Vocal fold posturing} \label{SecIIIA}
\begin{figure}[t] 
 \includegraphics{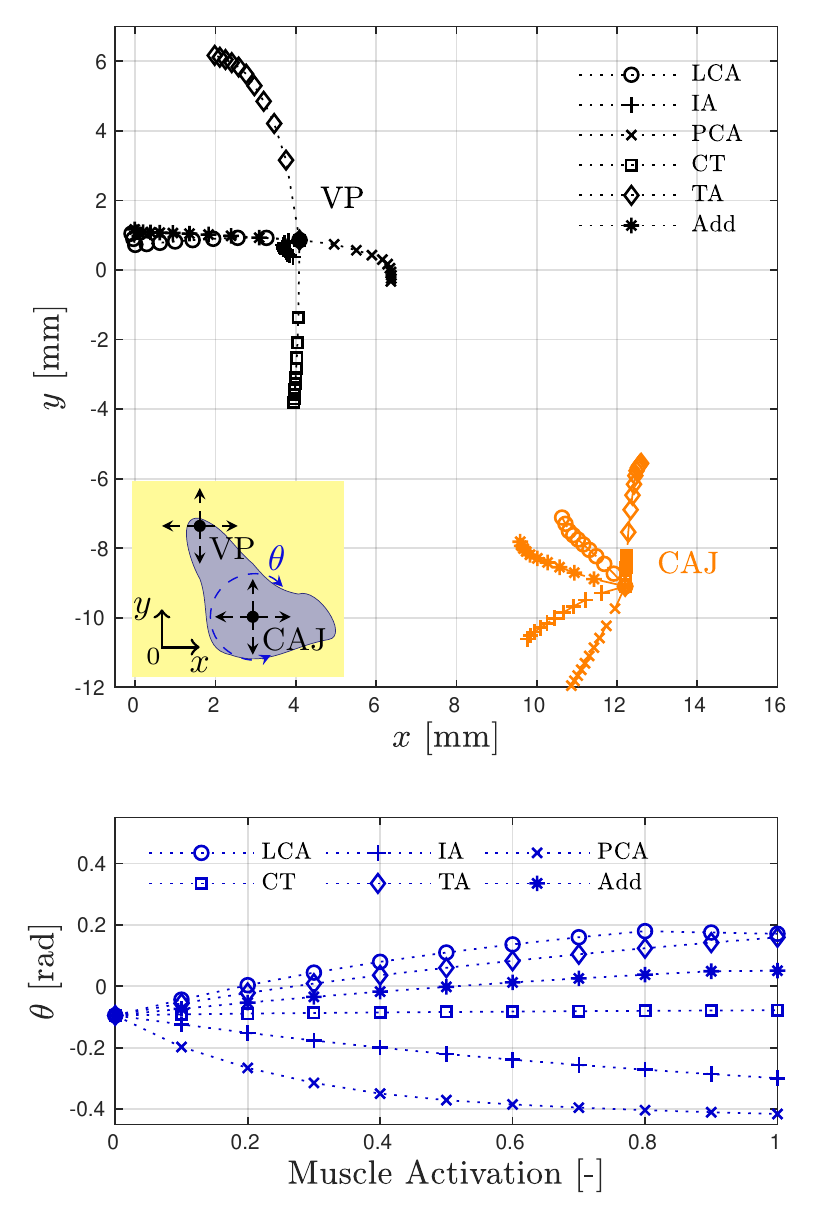}
 \caption{\label{FIG3} (Color online)
 Simulated accommodation of the (right) arytenoid cartilage obtained by the independent activation of the five intrinsic muscles and the adductory complex. \emph{Top:} Cartesian displacement of the cricoarytenoid joint (CAJ) center, $(x_\text{CAJ}+\xi, \, y_\text{CAJ}+\psi)$, and the vocal process (VP), $(x_{02}, \, y_{02})$.
 The inset schematic illustrates the VP movements resulting from the displacement and rotation of the CAJ.
 \emph{Bottom:} Rotation angle $\theta$ for the CAJ. Concurrent beginning of the paths indicates zero muscle activation.
 } 
\end{figure}

The biomechanical scheme for the prephonatory laryngeal posturing utilizes, principally, the arytenoid cartilage accommodation for controlling glottal adduction and abduction. \autoref{FIG3} shows the effect of the individual activation of the five intrinsic muscles on the (right) arytenoid posture, as described by the position of the cricoarytenoid joint and the vocal process. \autoref{FIG3} also includes an equivalent LCA/IA \emph{adductory} complex given by $a_\text{LCA} = a_\text{IA} = a_\text{Add}$, i.e., the coupled 1:1 activation for LCA and IA muscles, as in previous studies \citep{chhetri_neuromuscular_2012, chhetri_influence_2014, geng_threedimensional_2020, palaparthi_mapping_2019}. The muscle activation is incremented in normalized steps of $0.1$ from $0$ to $1$ for each case. Note that the displacement of the cricoarytenoid joint is smaller and different in nature than those of the vocal process. 

Herein, VF adduction is characterized by the medialization and (positive) counter-clockwise rotation of the arytenoid cartilage, and is controlled by the LCA muscle and, to a lesser extent, the TA muscle. Glottal abduction is achieved via the PCA muscle where concurrent vocal process lateralization and clockwise arytenoid rotation is observed. Furthermore, anterior-posterior displacements of the vocal process are determined by the antagonist effects of the TA and CT muscles. Although activation of the IA muscle minimally alters the position of the vocal process, it does displace the cricoarytenoid joint medially and caudally, thus reducing the posterior glottal opening. The adductory complex combines both the LCA and IA muscles, thus allowing for jointly reducing the membranous and posterior glottal areas.

\begin{figure}[t]
 \includegraphics{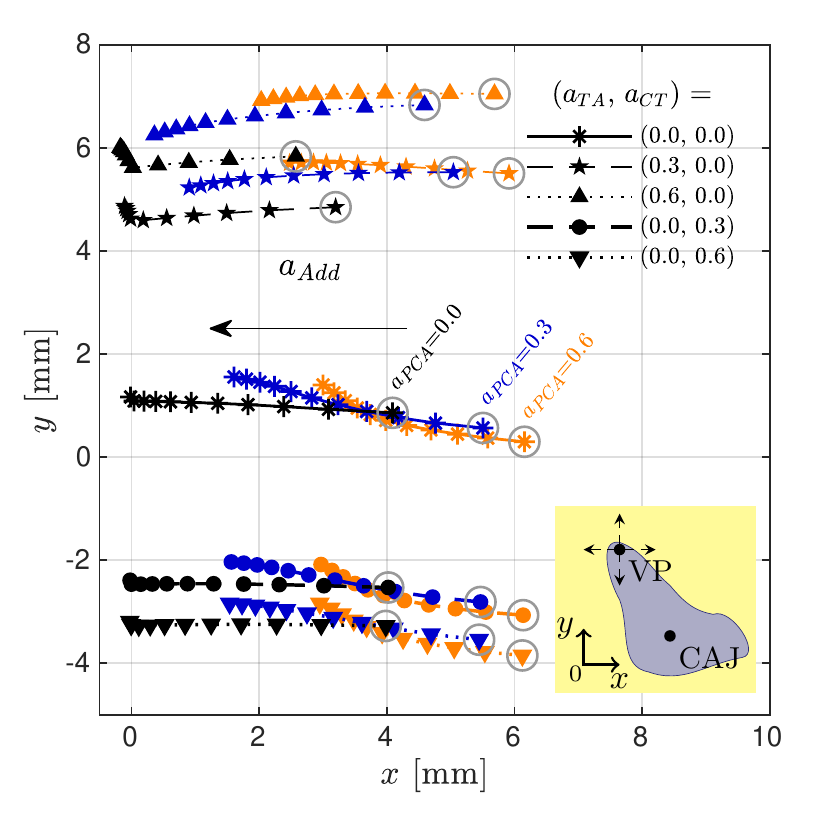}
 \caption{\label{FIG4} (Color online)
Effects of laryngeal muscle coactivation on glottal adduction. Vocal process (VP) coordinates, $(x_{02}, \, y_{02})$, produced through the activation of the adductory complex for the non-coactivation case (solid line with medium markers), and for two activation levels for both TA muscle (dashed and dotted lines with small markers) and CT muscle (dashed and dotted lines with large markers). Three PCA activations are drawn in black ($a_\text{PCA} = 0.0$), dark color ($a_\text{PCA} = 0.3$), and light color ($a_\text{PCA} = 0.6$) lines.
Markers enclosed in circles indicate the results for null adductory complex activation ($a_{Add}=0.0$). The inset schematic illustrates the VP Cartesian movements.
 } 
\end{figure}

The effects on VF adduction due to muscle coactivation in the model are illustrated in \autoref{FIG4}. The resulting movement of the vocal process (refer to the attached schematic) from the incremental activation of the adductory complex is shown for different coactivation states of the remaining muscles. For each path, activation levels for $a_\text{Add}$ increased from $0$ to $1$ in normalized steps of $0.1$. The resulting movement of the vocal process with no coactivation is drawn in black in the center of the figure, whereas the two topmost black paths and the two bottommost black paths describe the coactivation with the TA and CT muscles, respectively. Results involving two cases of PCA coactivation are also included for contrast. 

These results further illustrate the antagonist role of the TA and CT muscles, where ventral and dorsal displacements of the vocal process are produced via the activation of the TA and CT muscles, respectively, and where the TA muscle increases the total adduction of the vocal process. Moreover, \autoref{FIG4} also illustrates that coactivation of the PCA muscle introduces a noticeable lateral displacement in the vocal process path with increasing activation, and slightly changes the direction of its path. 

\begin{figure}[t]
 \includegraphics{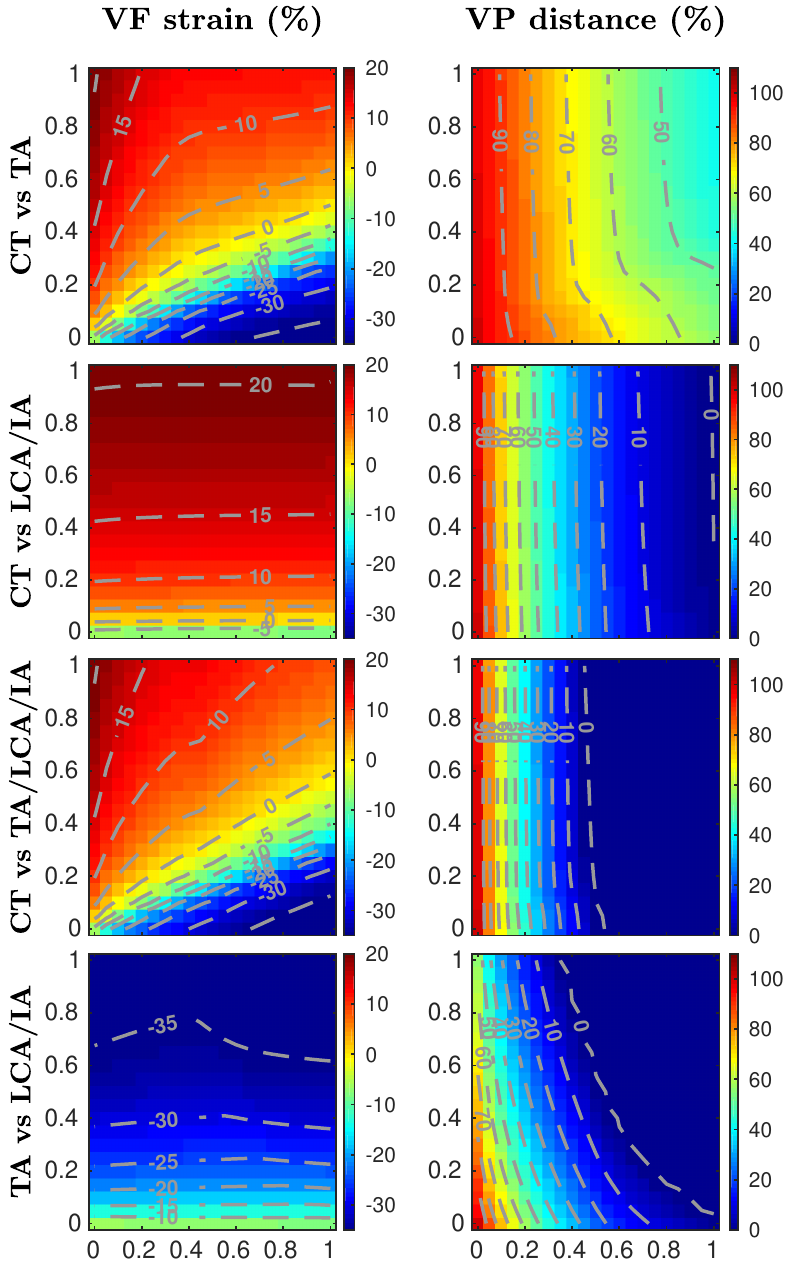}
 \caption{\label{FIG5} (Color online) Muscle activation plots for parametric coactivation of main phonatory intrinsic muscles. Vocal fold strain (left column) and vocal process distance (right column) are depicted as functions of the paired coactivation of muscle groups. The rows show four activation scenarios with contour lines included for clarity. For each row, A vs B refers to the abscissa and ordinate axes indicating activation levels for muscle groups A and B, respectively; null activations are set for the remaining intrinsic muscles.
 } 
\end{figure}

In \autoref{FIG5}, muscle activation maps depicting VF elongation and vocal process distance for pairs of intrinsic muscle groups are presented. VF elongation is computed by Equation \eqref{Eq:S2s1_09}, whereas the distance between the vocal processes is measured with respect to the rest distance at zero muscle activation.
A coupled 1:1 activation was applied for every muscle in a group, for example, an activation level $\bar{a}$ set to group LCA/IA implies $a_{LCA} = a_{IA} = \bar{a}$.
The coactivation of the main phonatory laryngeal muscles is investigated in \autoref{FIG5}, whereas the effect of PCA activation is described further bellow.
Similar representations have been applied in the past for investigating muscle groups having antagonistic or synergistic functions in VF posturing in experiments with excised canine larynges, \citep{chhetri_neuromuscular_2012, chhetri_influence_2014}, and in simulations involving a three-dimensional finite element model of a canine larynx \citep{geng_threedimensional_2020}. Given that the proposed biomechanical model of the larynx was partially based on canine models \citep{titze_twodimensional_2007}, contrast against these prior studies is feasible and meaningful.

Results for VF elongation (from VF strain in the left column of \autoref{FIG5}) show that the proposed model captures the antagonistic effects of the CT and TA muscles, where VF strain can be increased/decreased through corresponding CT/TA muscle activations. It can be observed that the main adductor muscles have a minor effect on VF elongation. The largest VF elongation (around +20\%) is obtained from maximum CT activation and zero TA activation. Similar strain values were obtained in simulations using finite element models \citep{geng_threedimensional_2020}, whereas studies with canine models have reported larger elongation values \citep{chhetri_neuromuscular_2012, chhetri_influence_2014}. In contrast, the shortest VF elongation results from maximum TA activation and zero CT activation. The minimum VF strain values (around -35\%) are lower than those reported in the benchmark studies. However, the strain contour lines for all activation conditions resemble the diagonal patterns reported by other authors \citep{chhetri_neuromuscular_2012,geng_threedimensional_2020}. 

Vocal process distance indicates the adduction of the vocal folds. The plots in \autoref{FIG5} show the resulting values for the  distance between the vocal processes, measured in percentage relative to 
the distance at rest (i.e., null activation for all intrinsic muscles). As expected, a strong synergistic adductory role played by the TA, LCA, and IA muscles is noticeable. Individual TA activation has a less significant effect on glottal adduction than the coactivation of the LCA/IA complex. However, the equal activation of the TA, LCA, and IA muscles as a group compresses the previous gradual posturing of the vocal processes at the region of lower activation levels. Despite the considerable anatomical and functional oversimplifications in the biomechanical model of the larynx, our simulations of vocal process distance are in agreement with previous experimental and high-order modeling studies \citep{chhetri_neuromuscular_2012, chhetri_influence_2014, geng_threedimensional_2020}. 
Two important differences are worth noting. First, the previous studies indicate that a non-zero vocal process distance is obtained even for strong cases of glottal adduction \citep{geng_threedimensional_2020, moisik_quantal_2017}. The implemented model, however, produces zero distance values for a range of muscle activation conditions. In addition, the current laryngeal model is not capable of reproducing the abductory effect of the CT muscle on VF adduction reported by other authors  \citep{chhetri_neuromuscular_2012, chhetri_influence_2014, geng_threedimensional_2020}. 

\begin{figure}[t]
 \includegraphics{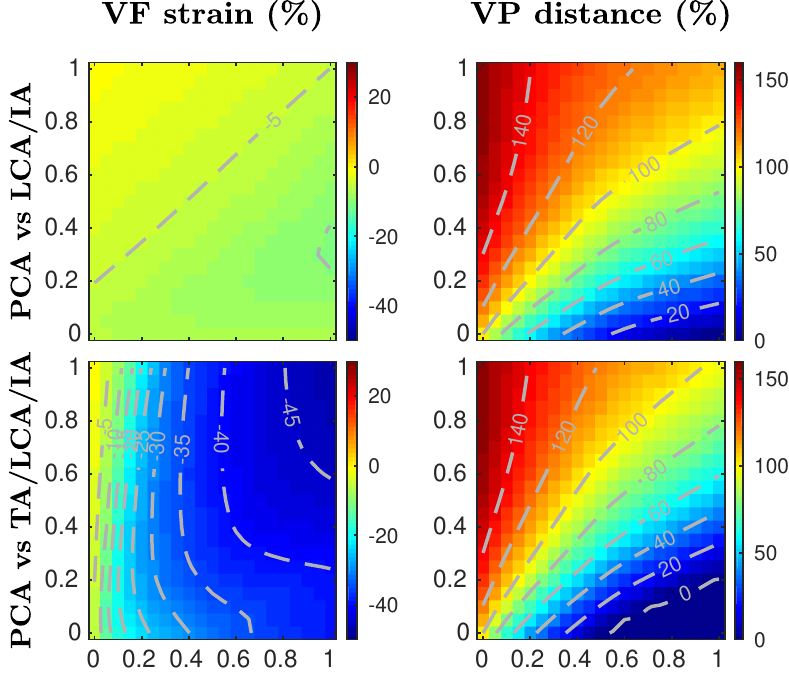}
 \caption{\label{FIG6} 
 (Color online) Muscle activation plots for parametric coactivation of adductory/abductory intrinsic muscles. Vocal fold strain (left column) and vocal process distance (right column) are depicted as functions of the paired coactivation of muscle groups. The rows show two activation scenarios with contour lines included for clarity. For each row, A vs B refers to the abscissa and ordinate axes indicating activation levels for muscle groups A and B, respectively; null activations are set for the remaining intrinsic muscles.
 } 
\end{figure}

The PCA muscle is the main larynx abductor, with a key role during inspiratory laryngeal movements, whereas during typical phonation, it plays only a minor role \citep{poletto_correspondence_2004}. Muscle activation plots portraying the VF elongation and vocal process distance for adductory/abductory antagonist muscle groups are depicted in \autoref{FIG6}, following the above methodology. The first row shows the results from the coactivation of the PCA muscle versus LCA/IA group. It is observed that the distance between the vocal processes ranges from nearly 0\% for full activation of the LCA/IA group and no activation of the PCA muscle, to around 150\% for the opposite case, whereas the changes in VF strain are minimal. Notice that, in the case of equal activation, the abductor force from the PCA muscle is slightly superior to the antagonist force produced by combining the LCA and IA muscles. The second row illustrates the results obtained by including the TA muscle in the adductor group. The results show that the TA muscle acts to reinforce the adductor forces that are opposite to the PCA muscle action, producing an appreciable region with null vocal process distance; moreover, TA muscle inclusion yields a dramatic VF shortening of up to -45 \% strain for full activation of the four muscles considered. Further comparisons depict that the PCA and CT muscles actuate differently during larynx accommodation, providing evidence of no antagonist effect; the PCA muscle abducts the vocal process producing no appreciable change in VF strain, whereas the CT muscle mainly elongates the VF with a minimal modification of the distance between the vocal processes. This results are not included here for conciseness.

\subsection{Sustained phonatory gestures} \label{SecIIIB}
Sustained vowel simulations are performed to investigate basic glottal function (self-sustained oscillation and pitch) as a function of the laryngeal muscle activation and subglottal pressure, $P_s$. For simplicity, the prephonatory laryngeal posture is fixed through the constant activations $a_\text{LCA} = a_\text{IA}=0.5$ and $a_\text{PCA}=0.0$ for the main adductor/abductor muscles, whereas VF adjustment is controlled via the parametric activation of $a_\text{CT}$ and $a_\text{TA}$ in the range from $0$ to $1$. 

\begin{figure*}[t]
 \includegraphics[width=16cm]{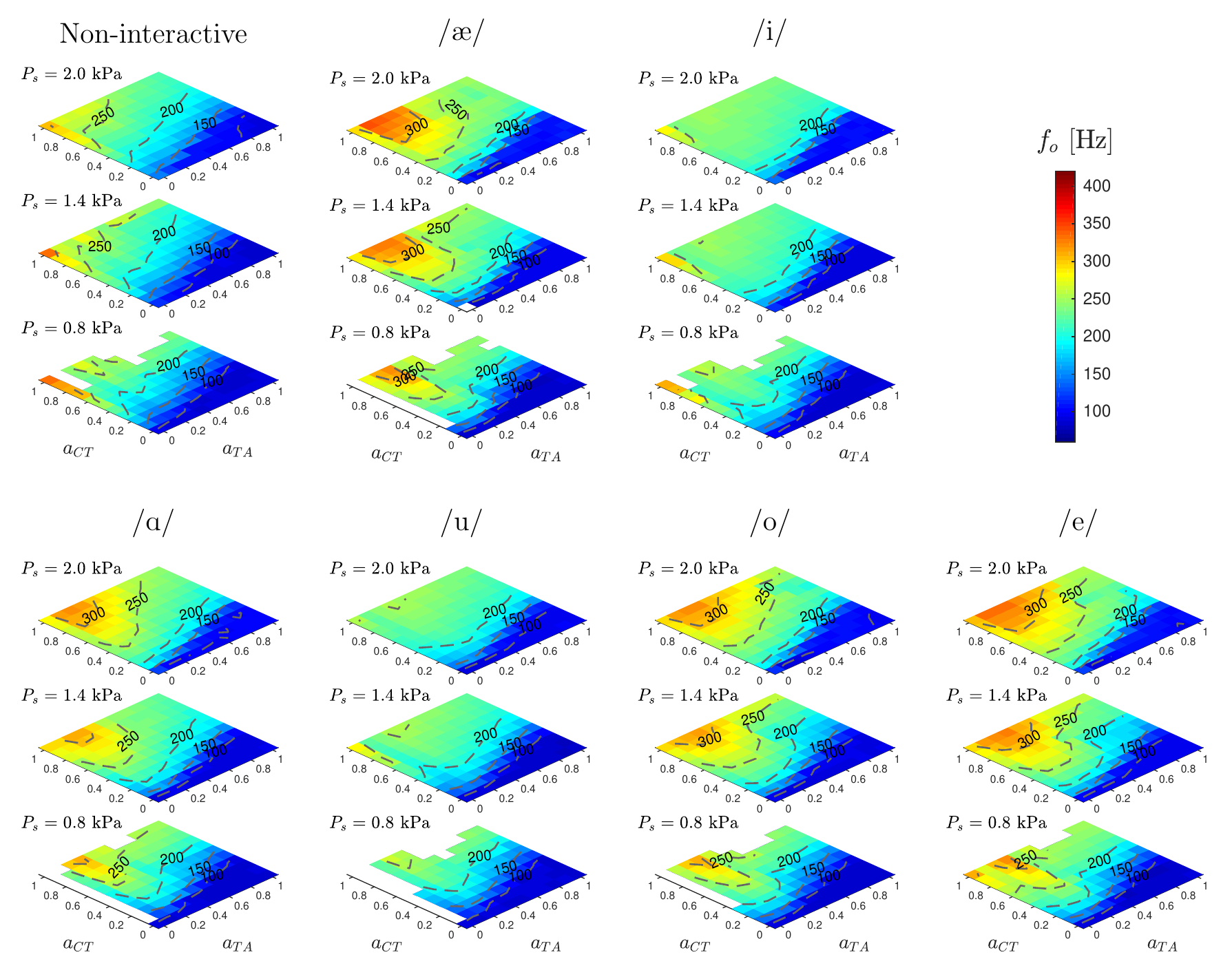}
 \caption{\label{FIG7} (Color online) Muscle activation plots with fundamental frequency for CT versus TA activation, considering different vocal tract shapes and subglottal pressures $P_s = [0.8, \, 1.4, \, 2.0]$ kPa. In all the cases $a_\text{LCA} = a_\text{IA}= 0.5$, and $a_\text{PCA}=0.0$. The non-interactive case has no vocal tract and represents an excised larynx scenario. Isofrequency contours are drawn for clarity.}
\end{figure*}

For each condition, 600 ms of simulated phonation data were obtained. In general, the simulations evidenced transients and fluctuations at the beginning, which were not relevant for the present analysis; thus, the first portion of data was discarded and only the last 200 ms steady segment was considered. An autocorrelation method was applied for assessing periodicity and fundamental frequency for the simulated glottal area waveform, $A_g$ \citep{mehta_using_2015}. In \autoref{FIG7}, muscle activation plots are shown for fundamental frequency as a function of TA and CT muscle coactivation, for three physiological subglottal pressure levels $P_s = [0.8, \, 1.4, \, 2.0]$ kPa. Acoustic coupling for vowels \{\textipa{/\ae/}, \textipa{/i/}, \textipa{/a/}, \textipa{/u/}, \textipa{/o/}, \textipa{/e/}\} as well as for a non-interactive acoustic system was also considered to further assess the model behavior. Results show that in the low subglottal pressure condition increasing CT activation leads to less stable, or even unachievable, VF oscillations (shown as blank portions in \autoref{FIG7}). Self-sustained oscillations are strengthened by increasing the subglottal pressure level. The relation between high CT activation and phonation onset pressure has been described in excised larynx experiments \citep{chhetri_influence_2014}, as well as in high-order simulations \citep{palaparthi_mapping_2019}. Results also show that the acoustic coupling facilitates self-sustained VF oscillations and raises pitch in comparison with the no vocal tract (non-interactive) scenario, where an increased region without periodic vibrations was observed. 

Fundamental frequency contours resulting from coactivation of the TA and CT muscles globally resemble those reported in \citet{titze_rules_2002} and \citet{lowell_simulated_2006}. CT activation is the main contributor to changes in fundamental frequency and the highest values are obtained for low values of TA activation. This result was expected and it has been reported before \citep{lowell_simulated_2006, titze_rules_2002}. However, the original rules for muscle control of the body-cover VF model yield fundamental frequencies greater than 500 Hz, whereas the frequency range in \autoref{FIG7} is more restricted, showing maximum values around 350 Hz. Fundamental frequencies up to 400 Hz are obtained for scenarios with increased gottal adduction. In addition, prior reported muscle activation plots featured a noticeable downward bending of the contour lines in the lower left region (low CT and TA). The implemented model also shows this non-linear response for the fundamental frequency in the region of weak TA activation and moderate-to-strong CT activation; however, bending in the contour lines are less pronounced in comparison with the results reported by \citet{titze_rules_2002}. This is more noticeable for the non-interactive vocal tract case.

\subsection{Dynamic phonatory gestures} \label{SecIIIC}
\begin{figure*}[t]
 \includegraphics[width=15cm]{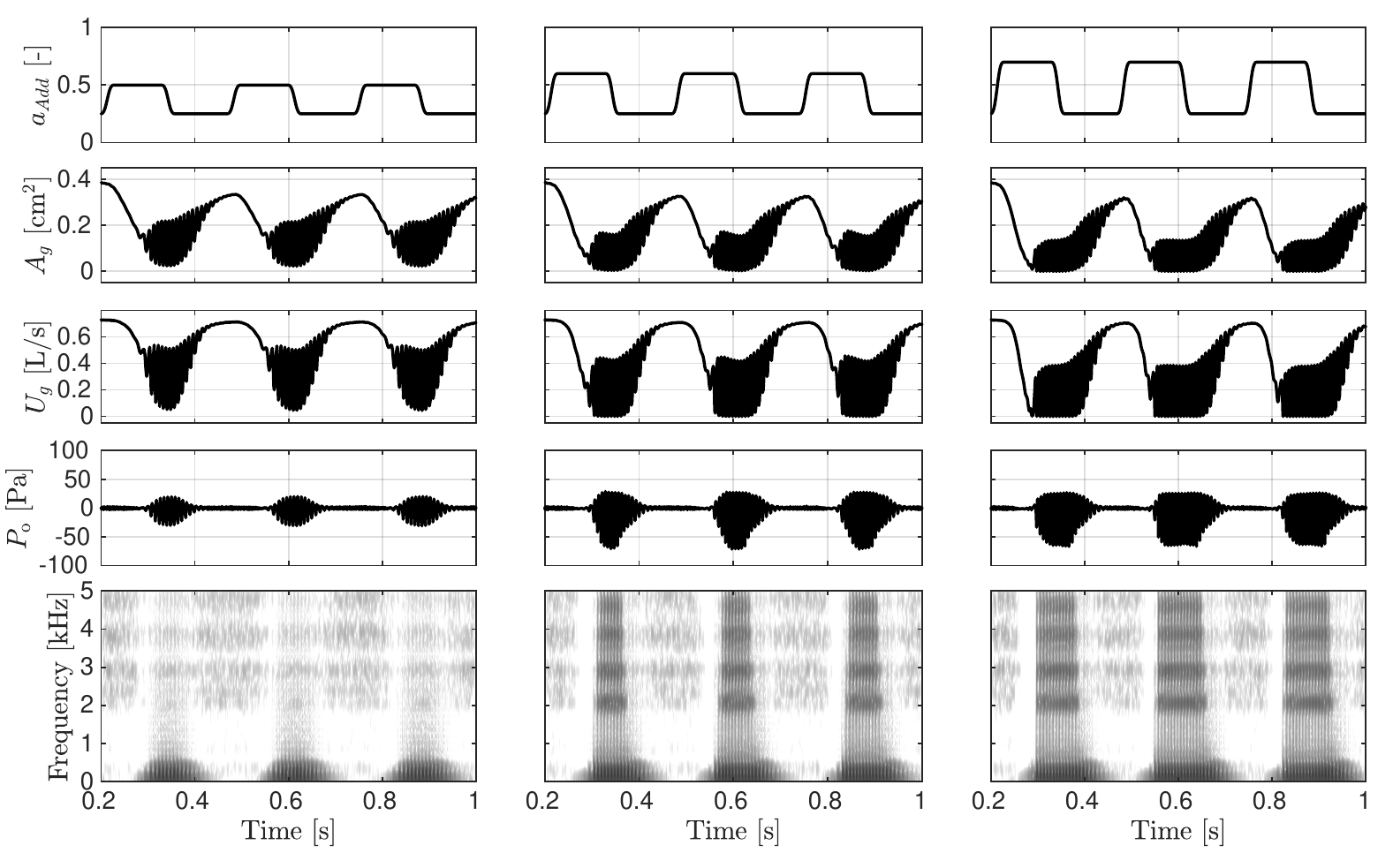}
 \caption{\label{FIG8} Simulated muscle control of voicing-devoicing for a \textipa{/hi-hi-hi-hi/} gesture. From left to right, columns correspond to simulations for weak, moderate, and strong VF adduction with maximum activation levels 0.5, 0.6, and 0.7, respectively. Adductory signals $a_\text{Add}$ are shown in the top row. The following three rows show the output signals for the glottal area $A_g$, glottal volume velocity $U_g$, and the radiated acoustic pressure $P_\mathrm{o}$. The bottom row depicts the wideband spectrogram of $P_\mathrm{o}$.
  }
\end{figure*}

Simulations of dynamic (articulatory) gestures combine posturing and phonation and are key to assessing the overall performance of the model. Simulations of repetitive voicing-devoicing gestures, i.e.,  \textipa{/hi-hi-hi-hi/}, are shown in \autoref{FIG8}. Model parameters are set to produce a modal male voice, where $P_{s}=800$ Pa, $a_\text{TA} = 0.2$, $a_\text{CT} = 0.1$, and $a_\text{PCA} = 0.0$, for a (male) vowel \textipa{/i/} for a direct comparison with \citet{titze_twodimensional_2007}. 

As before, the equivalent LCA/IA adductory complex is controlled in time with $a_\text{LCA}=a_\text{IA}=a_\text{Add}$. The three columns in \autoref{FIG8} show simulations for weak (left), moderate (middle), and strong (right) VF adduction, respectively. Top panels depict the muscle control signal $a_\text{Add}$ for every case, with peak amplitude indicating the adduction level. Waveform outputs of glottal area $A_g$, glottal volume velocity $U_g$, and radiated pressure $P_\mathrm{o}$ for the three simulated adduction strengths are shown in the following three rows, respectively. Spectrograms describing time-frequency content for the radiated pressure signals are included in the bottom row.

Despite the overly simple and fairly square activation signals, the simulated variables capture the gradual VF adduction/abduction characteristic of the voicing/devoicing gesture. The noticeable delay in area and airflow waveforms with respect to variations in the control signal, $a_\text{Add}$, is a product of the combined effects of inertia in the biomechanical model of the larynx and the time constants in the laryngeal tissue model. Radiated sound pressure is further delayed as a consequence of acoustic wave propagation throughout the vocal tract.

Changes in VF adduction are primarily seen in the voice onset time and, to a lesser extent, in the voice offset time. A higher magnitude in $a_\text{Add}$ speeds up the onset of VF oscillations and delays the offset of voicing. VF adduction condition also affects phonation time and the spectral richness of the radiated sound. The latter is in accordance with previous studies illustrating that, for a constant driving pressure, weak to moderate adduction conditions modify PGO area in the range of small to large, which in turn produces a considerable reduction of the radiated sound pressure level \citep{galindo_modeling_2017, zanartu_modeling_2014}.

Note that simulations are built around muscle control signals for the LCA and IA muscles only. However, it is well known that during speech production all the laryngeal muscles coactivate in complex ways \citep{moisik_quantal_2017, movahhedi_effects_2021, zhang_mechanics_2016}. As illustrated before, other muscles, such as PCA, TA, and CT, have an effect on VF adduction and can also alter voicing onset and offset \citep{poletto_correspondence_2004, titze_twodimensional_2007}. The proposed model allows for incorporating all five muscle activations in a time varying fashion, which can  improve the physiological relevance when modeling articulatory gestures. Nevertheless, the resulting output with the simplified muscle control input is still in agreement with \citet{titze_twodimensional_2007} and \citet{poletto_correspondence_2004}.

\subsection{Antagonistic muscle behavior}\label{SecIIID}

As the proposed model allows for analyzing the effects of selected levels of antagonistic muscle activation, we  herein simulate conditions that have the same vocal fold posturing, but with different underlying antagonistic muscle activations. We argue that this exercise can lead to future studies exploring disproportionate coactivation of antagonist intrinsic muscles that can be related to NPVH.

\begin{figure}[t]
 \includegraphics{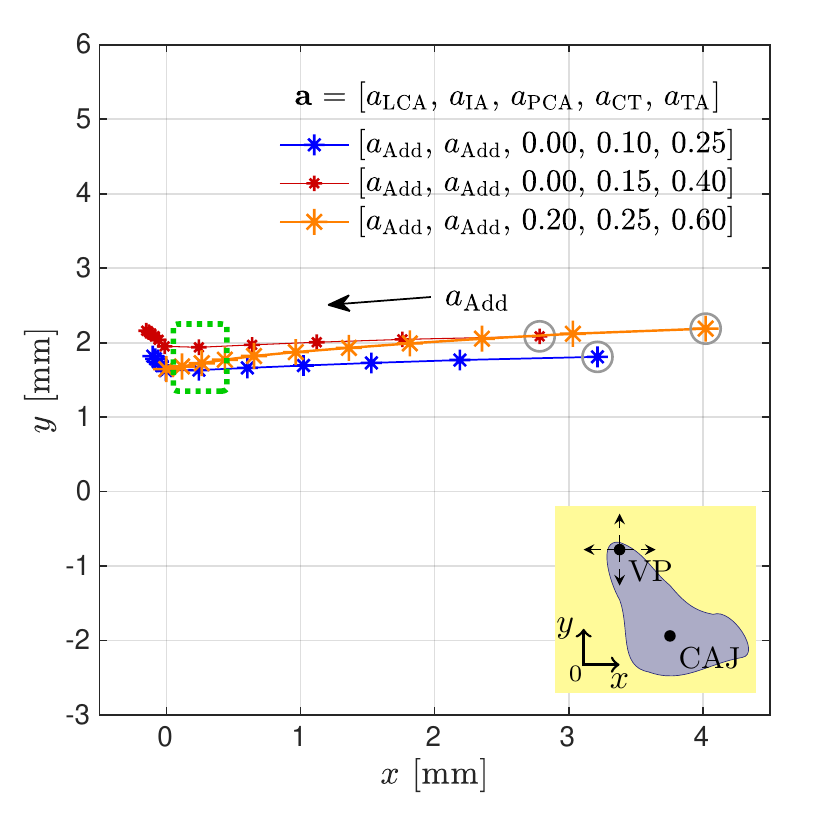}
 \caption{\label{FIG9} Displacement paths for the right vocal process (VP), $(x_{02}, \, y_{02})$, for three simulated tension states in the larynx. Trajectories correspond to laryngeal postures produced by parametric activation of the intrinsic muscles following the activation set $\mathbf{a} = [a_\text{LCA}, \,a_\text{IA}, \,a_\text{PCA}, \,a_\text{CT}, \,a_\text{TA}]$. The paths describe variations for the normalized activations
 $a_\text{Add}$ in the range of $0$ to $1$ in steps of $0.1$. 
The inset schematic illustrates the VP Cartesian movements. Configurations enclosed in the dotted rectangle are considered for voiced sound simulations.
  } 
\end{figure}

The ability of the model to attain the same prephonatory VF posturing for various muscle activation conditions is illustrated in \autoref{FIG9}, where the displacement paths for the right vocal process for three simulated laryngeal conditions are drawn. The dotted rectangle in \autoref{FIG9} shows that three distinctly different activation conditions (reported in \autoref{Tab2}) can produce almost the same vocal process position, and result in a visually similar VF configuration. This idea can also be extended into paths of glottal adduction. 
\autoref{FIG9} shows three paths obtained when varying the adductory complex $a_\text{Add}$ that correspond to different activations of all intrinsic muscles. These paths compare a given prephonatory state with two other conditions involving lower and higher activation levels, respectively. A slight increase in abduction is seen in the latter case for each marker (displacement to the right) do to PCA activation; although the paths overlap, this indicator shows that the increased tension scenario yields a more abducted condition for every step.

Even though VF posturing can be visually preserved, changes in muscle activation are expected to alter the TBCM parameters differently and, thus, the resulting vocal function. To investigate this idea, sustained phonation is simulated for the three activation sets (denoted as  $\mathbf{a}_{I}$, $\mathbf{a}_{II}$, and $\mathbf{a}_{III}$ in \autoref{Tab2}) that result in the same VF posture although associated with different degrees of coactivation of the antagonist intrinsic muscles. For these simulations, a driving pressure $P_{s} = 800$ Pa is applied, and tract area functions for ten different vowels. 

To describe multiple aspects of vocal function for each set, various parameters were computed, including 1) VF posturing: VF strain, vocal process distance, and membranous and posterior portions of the glottal area; 2) VF kinematics: fundamental frequency, open quotient (OQ), and closing quotient (ClQ); 3) Glottal aerodynamics: unsteady glottal airflow (AC Flow), stable glottal flow (DC Flow), maximum flow declination rate (MFDR), and amplitude of the first harmonic relative to the second harmonic ($H_1-H_2$); 4) Voice acoustics: sound pressure level (SPL), and low-to-high ($L/H$) spectral ratio. These measures have been shown to be valuable for the assessment of vocal hyperfunction \citep{galindo_modeling_2017, mehta_using_2015, zanartu_modeling_2014, Cortes2018, Mehta2019}. 

The resulting measures of vocal function are reported in \autoref{Tab3}. As expected, given that the vocal process position is almost the same, the parameters describing the laryngeal posture are similar for the three conditions; the only evident difference is the increased PGO area for set $\mathbf{a}_{II}$ that results from the lower activation of the LCA and IA muscles. Moreover, a significant increase in the minimum glottal area is observed for the $\mathbf{a}_{II}$ condition, which leads to a higher DC Flow. 
Kinematic measures in \autoref{Tab3} show that fundamental frequency and closing quotient show minor changes with the activation conditions, whereas OQ is noticeably increased for the $\mathbf{a}_{II}$ condition.

Aerodynamic measures in \autoref{Tab3} show that the highest MFDR values are obtained, as expected, for the more adducted cases, i.e., with less PGO leakage. Conversely, increasing the PGO area decreases MFDR, suggesting less efficient acoustic excitation. Interestingly, AC Flow and $H_1-H_2$ show minor variations, suggesting negligible differences in the vocal tract coupling and the source spectral tilt. In addition, acoustic measures provide complementary evidence of the reduced phonatory efficiency for set $\mathbf{a}_{II}$, yielding a noticeable decrease of SPL levels together with increased L/H spectral ratios, which are signs of less spectral energy in the high frequency region.

\section{Discussion} \label{Sec.Disc}

The proposed scheme allows for controlling prephonatory posturing and phonatory gestures in a triangular body-cover model of vocal folds by means of the independent activation of all of the intrinsic laryngeal muscles. It combines prior efforts to model laryngeal posturing \citep{titze_twodimensional_2007}, empirical rules relating muscle activation to biomechanical parameters in a lumped-element model \citep{titze_rules_2002}, and a triangular body-cover VF model \citep{galindo_modeling_2017}. The scheme thus provides a flexible and physiologically relevant way to control the self-sustained fully interactive voice production model for both sustained vowels and time-varying glottal gestures. As a result, the approach allows for exploring the role of antagonistic muscle pairs in phonation, which has direct implications for studying normal and disordered muscle behavior in phonation. 

In this study, it was first shown that the implemented framework can reproduce prior findings with excised larynx experiments and high-order numerical simulations for sustained vowels and simple articulatory gestures. In addition, we applied the approach to exemplify the role of antagonistic muscle pairs and the effects of antagonistic muscle activation in the context of a common manifestation of NPVH, with simulations in agreement with prior clinical data. Thus, despite its low-order complexity, the scheme provides a physiologically-inspired tool that could be used to gain insights into the physical mechanisms underlying intrinsic muscle function in phonation. This becomes particularly useful for extending previous efforts to model vocal hyperfunction \citep{zanartu_modeling_2014, galindo_modeling_2017} to account for elevated, unbalanced, and poorly regulated activity of the intrinsic laryngeal muscles, as seen in NPVH \citep{Hillman_framework_2020}. We hypothesize that NPVH is associated with a disproportionate coactivation of antagonist intrinsic muscles that result in a more abducted prephonatory VF posturing, that in turn results in a reduced loudness that is compensated with elevated subglottal pressure and further muscle tension. Subsequent efforts will explore the use of the proposed model to investigate these ideas and aid in delineating the etiology and pathophysiology of vocal hyperfunction.
The model has also potential for developing improved estimation frameworks based on biomechanical models of human phonation applicable to the prevention and treatment of VH  \citep{alzamendi_bayesian_2020,ibarra_estimation_2021}.

An evident limitation regarding the implemented muscle control scheme is worth highlighting. The effects of the coupled activation of the five intrinsic muscles on the glottal posture and vocal fold configuration are simulated by applying the biomechanical model by \citet{titze_twodimensional_2007} at a low-to-moderate computational cost. However, this model is built around the effective displacement and rotation of the laryngeal cartilages on a transverse glottal plane, and is not thus able to describe in full detail the 3D glottal adjustment, in contrast with, for example, the anatomically accurate finite element models. Current evidence supports that the arytenoid movement relative to an axis located at the CAJ and perpendicular to the transverse glottal plane (see \autoref{FIG1a}) is highly constrained, and that a rocking-sliding motion about the long axis of the cricoid cartilage is more accurate \citep{ geng_threedimensional_2020, yin_interaction_2014}. The latter can explain in part the differences in the simulated vocal fold posturing discussed in \autoref{SecIIIA}, and it is expected to also affect the three way coupling between tissue, airflow, and sound at the glottis. Furthermore, the 2D scheme neglects the changes in the medial surface shape and the vertical height of the superior edge of the vocal folds \citep{movahhedi_effects_2021}, and underestimates the anatomical constraints with extrinsic structures which can play a role during phonation \citep{moisik_quantal_2017}.

Two important aspects regarding the biomechanical modeling of the laryngeal tissues for the modified Kelvin model are highlighted. First, model parameters selection is key for producing reliable responses. Although a number of references report biomechanical parameters for the main laryngeal tissues in humans \citep{hunter_threedimensional_2004, hunter_refinements_2007, palaparthi_mapping_2019, titze_myoelastic_2006}, the significant variability among the different reported values is striking. The lack of consistent human data required a compromise that considered laryngeal function in both animals and humans for the model. Thus, the ad hoc values reported in \autoref{TabA1} correspond to those tissue parameters that yield the best performance simultaneously for both the laryngeal posture model and the TBCM; notwithstanding, significant differences with prior works are obtained, as evidenced in the simulated muscle activation plots with vocal process distance and fundamental frequency. In addition, each laryngeal tissue was modeled as a single one-dimensional fiber. This is a limitation to accurately capture the distributed portions of the intrinsic muscles, especially for the thyrovocalis and thyromuscularis portions of the TA muscle. This could be extended by adding independent submodels representing the different muscle portions, at the expense of increasing the model complexity. We opted to keep the model with the single one-dimensional fiber assumption for simplicity. 

Note that the proposed scheme for the TBCM model continues to be an approximation and has limitations that could be addressed in future studies, such as constraints for cartilage displacements \citep{hunter_threedimensional_2004, geng_threedimensional_2020}, superior-inferior accommodation of the larynx during phonation \citep{moisik_quantal_2017}, medial bulging, and anterior-posterior gradient, among others.

Further assessment using heuristic or random control rules could provide additional insights into the muscle control of the laryngeal function to disentangle the relations between laryngeal function and the resulting acoustic output. Other future avenues for exploration include the connection with a neurophysiological muscle activation that incorporates natural neurological fluctuations in the activation of intrinsic laryngeal muscles \citep{manriquez_neurophysiological_2019}, and simulations of /VCV/ gestures to compute relative fundamental frequency.

\section{Conclusions} \label{Sec.Concl}

A physiological scheme for controlling the mechanical properties of a triangular body cover model of the vocal folds through the independent activation of the five intrinsic laryngeal muscles is proposed. The approach builds upon prior efforts that describe rules for controlling low-order models, vocal fold posturing, and a triangular vocal fold model. The scheme provides a flexible and physiologically relevant way to control the self-sustained fully interactive voice production model for both sustained vowels and time-varying glottal gestures. At the same time, the resulting model allows for exploring the role of antagonistic muscle pairs in phonation. The model simulations are in agreement with prior studies using excised larynx experiments and high-order simulations. Using the proposed approach, we illustrate that different states of activation can lead to the same vocal fold posturing, albeit with highly different stress states. These similar posturing scenarios, however, have clear differences in the resulting kinematic, aerodynamic, and acoustic measures of vocal function. The resulting model is a relevant tool that can provide key insights into the physical mechanisms underlying normal and disordered phonation. 

\begin{acknowledgments}
This work was supported by ANID grants FONDECYT 1191369 and BASAL FB0008, STIC AmSud ASPMLM-Voice 21-STIC-05, and the National Institute on Deafness and Other Communication Disorders of the National Institutes of Health under award number P50DC015446. The content is solely the responsibility of the authors and does not necessarily represent the official views of the National Institutes of Health.
\end{acknowledgments}

\appendix

\section{Modified Kelvin model} \label{Sec.KelvinModel}
The modified Kelvin model applied for simulating the laryngeal tissue is briefly described. It is a one-dimensional,  biomechanical model for the internal stress-strain response in fibrous tissue given by Equation \eqref{Eq:A1_01}.
The main simulated variables are the axial stress, $\sigma_{i}$, and the axial strain, $\epsilon_{i}$, for each tissue $i \in \mathcal{I}$. Equation \eqref{Eq:A1_01a} represents the total (passive plus active) stress, whereas Equation \eqref{Eq:A1_01b} describes active stress due to the internal contractile properties in the intrinsic muscles.
\begin{subequations}
  \label{Eq:A1_01}
\begin{eqnarray}
  \label{Eq:A1_01a}
      t_{s} \dot{\sigma}_{i} + \sigma_{i} & = & \left[ \sigma_{a\,i} + \sigma_{p\,i} + E t_{p} \dot{\epsilon_{i}} \right], \\
  \label{Eq:A1_01b}
      t_{a} \dot{\sigma}_{a\,i} + \sigma_{a\,i} & = & a_{i} \sigma_{m} \max \left\{0, \, 1 - b (\epsilon_{i} - \epsilon_{m})^2 \right\},
\end{eqnarray}
\end{subequations}
where a dot over a variable indicates time derivative, $\sigma_{a\,i}$ is the active stress, and $\sigma_{p\,i}$ is the passive viscoelastic stress corresponding to the fiber deformation $\epsilon_{i}$ modeled as follows
\citep{hunter_threedimensional_2004,titze_myoelastic_2006}:
\begin{equation}
  \label{Eq:A1_02}
  \begin{aligned}
  \sigma_{p\,i} & = -\frac{\sigma_{0}}{\epsilon_{1}} (\epsilon_{i} - \epsilon_{1}), & \epsilon_{i} \leq \epsilon_{2},  \\
                   & = -\frac{\sigma_{0}}{\epsilon_{1}} (\epsilon_{i} - \epsilon_{1})
                       + \sigma_{2} \left[e^{B (\epsilon_{i} - \epsilon_{2})} -1 
                       - B (\epsilon_{i} - \epsilon_{2}) \right],  & \epsilon_{i} > \epsilon_{2}.
  \end{aligned}
\end{equation}
For non-contractile tissue (i.e., vocal ligament and mucosa) the active component is set to zero.

Tissue-specific dynamical properties and stress-strain response characteristics in the Kelvin model of Equation \eqref{Eq:A1_01} can be specified independently by its parameters \citep{hunter_threedimensional_2004, titze_myoelastic_2006}: $t_{s}$ is a time-series constant, $t_{p}$ is a parallel time constant, $t_{a}$ is the internal activation time constant, $A_{i}$ is the cross-sectional area, $\sigma_{m}$ is the maximum isometric active stress, $\epsilon_{m}$ is the strain at maximum contractile stress, $b$ is a coefficient for active stress, $\sigma_{0}$ is the stress at zero strain, $\sigma_{2}$ is a scale factor for the exponential function, $B$ is an exponential strain constant, $\epsilon_{1}$ is the strain at zero stress, and $\epsilon_{2}$ is the strain at which the nonlinear exponential function begins. Moreover, $E = \frac{d \sigma_{p\,i}}{d \epsilon_{i}}$ is the nonlinear tangent Young's modulus. The resulting fiber force magnitude is $F_{i} = A_{i} \sigma_{i}$, where the force direction coincides with the longitudinal axis in the fiber. 
The parameters set in this work for modeling laryngeal tissue are reported in \autoref{TabA1}, whereas the remaining parameters are set as originally introduced in \citep{titze_myoelastic_2006}.  




\begin{table*}[p]
 \caption{\label{Tab1}
 Anatomical and biomechanical parameters required for implementing the dynamic simulation of the laryngeal posturing and glottis configuration.}
 \begin{ruledtabular}
 \begin{tabular}{lcc}
 & Definition & Value \\
 \hline
 \multirow{4}{*}{CAJ} &
 Mass and moment of inertia &
 $M_\text{ac} = 1.4 \times 10^{-3}$ kg, $I_\text{ac} = 1.6\times 10^{-6}$ kg m$^2$. \\
 & Vocal process cadaveric position & $\bar{x}_{0} = 4$ mm, $\bar{y}_{0} = 0$ mm. \\
 & CAJ center coordinates & $x_\text{CAJ} = 10.1$ mm, $y_\text{CAJ} = -10.1$ mm. \\ 
 & Translational/rotational dampings & $d_{x}= 0.02$ s, $d_{y}=0.02$ s, $\delta=0.02$ s. \\
 \hline
 \multirow{4}{*}{CTJ} &
 Mass and moment of inertia &
 $M_\text{cc} = 10^{-2}$ kg, $I_\text{cc} = 10^{-5}$ kg m$^2$. \\
 & CT/TA moment arms &
 $r_\text{TA} = 16.1$ mm, $r_\text{CT} = 11.1$ mm. \\
 & CT angle relative to TA &  $\phi = 45^{\circ}$ , $\cos \phi = 0.76$. \\
 & Translational/rotational viscous times & $t_{t} = 0.04$ s, $t_{r} = 0.04$ s. \\
 \hline
 \multirow{2}{*}{Glottis} &
 Rest VF length & $L_0 = 16$ mm. \\
 & Posterior wall half-width coordinates & $x_{p2} = 3.2$ mm, $y_{p} = -2.5$ mm. \\
 \end{tabular}
 \end{ruledtabular}
 \end{table*}



\begin{table}[p]
 \caption{\label{Tab2}
   Three sets of appreciable different intrinsic muscle activation levels producing nearly the same simulated vocal process configurations. 
 }
 \begin{ruledtabular}
 \begin{tabular}{ccccccc}
      Set & $a_{LCA}$ & $a_{IA}$ & $a_{PCA}$ & $a_{CT}$ & $a_{TA}$  \\ \hline 
      $\mathbf{a}_{I}$ & 0.50 & 0.50 & 0.00 & 0.10 & 0.25 \\
      $\mathbf{a}_{II}$ & 0.40 & 0.40 & 0.00 & 0.15 & 0.40 \\
      $\mathbf{a}_{III}$ & 0.80 & 0.80 & 0.20 & 0.25 & 0.60 \\
 \end{tabular}
 \end{ruledtabular}
\end{table}



\begin{table*}[t]
 \caption{\label{Tab3} Effects of three different activation conditions on sustained phonation. Reported prephonatory posture parameters were obtained from the biomechanical larynx model. Kinematic, aerodynamic, and acoustic mean (SD) parameters were computed from simulated phonatory data considering tract area functions for 
 vowels [\textipa{\ae} \textipa{2} \textipa{A} \textipa{e} \textipa{E} \textipa{I} \textipa{i} \textipa{o} \textipa{u} \textipa{U}] 
 and driving pressure $P_{s} = 800$ Pa.
 }
 \begin{ruledtabular}
 \begin{tabular}{ccccc} 
  & & \multicolumn{3}{c}{Activation sets} \\ 
  \cline{3-5} 
  & & $\mathbf{a}_{I}$ & $\mathbf{a}_{II}$ & $\mathbf{a}_{III}$ \\ 
  \hline 
  \multirow{4}{*}{\textbf{Posture}} 
  & Strain $\epsilon$ [\%] & -9.5 & -11.3 & -10.3 \\ 
  & VP Distance [\%] & 5.3 & 6.0 & 6.3 \\ 
  & $A_{MGO}$ [mm$^2$] & 0.8 & 0.9 & 1.0 \\ 
  & $A_{PGO}$ [mm$^2$] & 1.4 & 3.0 & 1.1 \\ 
  \hline 
  \multirow{3}{*}{\textbf{Kinematic}} 
  & $f_{0}$ [Hz] & 116.4 (4.7) & 105.1 (4.4) & 115.1 (3.7) \\ 
  & OQ [\%] & 81.7 (7.6) & 98.6 (3.8) & 79.7 (11.7) \\ 
  & ClQ [\%] & 24.0 (8.5) & 33.9 (5.9) & 25.5 (8.4) \\ 
  \hline 
  \multirow{4}{*}{\textbf{Aerodynamic}} 
  & AC Flow [mL/s] & 395.2 (34.1) & 367.6 (35.0) & 372.2 (28.1) \\ 
  & DC Flow [mL/s] & 11.1 (1.1) & 65.6 (1.7) & 6.1 (1.0) \\ 
  & MFDR [L/s$^2$] & 710.6 (146.9) & 375.7 (92.7) & 648.4 (133.2) \\ 
  & H1-H2 [dB] & 12.9 (1.5) & 12.7 (1.1) & 12.2 (1.4) \\ 
  \hline 
  \multirow{2}{*}{\textbf{Acoustic}} 
  & SPL [dB] & 84.9 (3.5) & 77.2 (3.5) & 84.5 (3.4) \\ 
  & L/H Ratio [dB] & 31.0 (14.1) & 41.4 (4.7) & 31.7 (14.1) \\ 
\end{tabular} 
\end{ruledtabular}
\end{table*}



\begin{table*}[t]
 \caption{\label{TabA1}
 Parameters for simulating laryngeal tissues according to the modified Kelvin model. Passive stress-strain response and active stress for the five intrinsic laryngeal muscles -CT: cricothyroid, TA: thyroarytenoid, LCA: lateral cricoarytenoid, IA: interarytenoid, and PCA: posterior cricoarytenoid-, the vocal ligament (LIG) and mucosa (MUC) are considered.}
 \begin{ruledtabular}
 \begin{tabular}{cccccccc}
 & \multicolumn{7}{c}{Laryngeal muscles and tissues} \\
 \cline{2-8}
     Parameter & CT & LCA & TA & IA & PCA & LIG & MUC \\ \hline 
   $\sigma_0$ [kPa] & 2.2 & 3.0 & 2.0 & 2.0 & 5.0 & 2.0  & 1.0 \\
   $\sigma_2$ [kPa] & 5.0 & 59.0 & 1.5 & 30.0 & 55.0 & 1.4 & 20.0 \\
   $B$ [-] & 7.0 & 4.0 & 6.5 & 3.5 & 5.3 & 13.0 & 4.4 \\
   $\epsilon_1$ [-] & -0.5 & -0.5 & -0.5 & -0.5 & -0.5 & -0.5 & -0.5 \\
   $\epsilon_2$ [-] & -0.06 & -0.06 & -0.05 & -0.06 & -0.05 & -0.3 & -0.3 \\
   $\sigma_m$ [kPa] & 300 & 100 & 150 & 100 & 100 & --- & --- \\
   $\epsilon_m$ [-] & 0.0 & 0.4 & 0.2 & 0.4 & 0.4 & --- & --- \\
 \end{tabular}
 \end{ruledtabular}
 \end{table*}


 \clearpage
 \listoffigures
 
\end{document}